\documentclass[universe,article,accept,pdftex,moreauthors]{Definitions/mdpi} 

\firstpage{1} 
\pubvolume{1}
\issuenum{1}
\articlenumber{0}
\pubyear{2024}
\copyrightyear{2024}
\externaleditor{Academic Editor: Firstname Lastname\vspace{-12pt}}
\datereceived{31 January 2024} 
\daterevised{18 March 2024} 
\dateaccepted{26 March 2024} 
\datepublished{ } 
\hreflink{https://doi.org/} 
\usepackage{tablefootnote}

\Title{New Timing Results of MSPs from NICER Observations}

\TitleCitation{New Timing Results of MSPs from NICER Observations}

\Author{Shijie Zheng $^{1,2}$*\orcidA{}, Dawei Han $^{1,2}$, Heng Xu $^{3}$, Kejia Lee $^{3,4}$, Jianping Yuan $^{2,5}$, Haoxi Wang $^{1,2}$, Mingyu Ge $^{1,2}$, \linebreak Liang Zhang $^{1,2}$,  Yongye Li $^{1,6}$, Yitao Yin $^{1,2}$, Xiang Ma $^{1,2}$, Yong Chen $^{1,2}$ and Shuangnan Zhang $^{1,2}$}

\AuthorNames{Shijie Zheng, Dawei Han, Heng Xu, Kejia Lee, Haoxi Wang, Mingyu Ge, Jianping Yuan, Liang Zhang, Yongye Li, Yitao Yin, Xiang Ma, Yong Chen, and Shuangnan Zhang}

\AuthorCitation{Zheng, S.; Han, D.; Xu, H.; Lee, K.; Wang, H.; Ge, M.; Yuan, J.; Zhang, L.; Li, Y.; Yin, Y.; et al.}

\address{%
$^{1}$ \quad Key Laboratory of Particle Astrophysics, Institute of High Energy Physics, Chinese Academy of Sciences, Beijing 100049, China; dwhan@ihep.ac.cn (D.H.); wanghaoxi@ihep.ac.cn (H.W.); gemy@ihep.ac.cn (M.G.); zhangliang@ihep.ac.cn (L.Z.); yyli@ihep.ac.cn (Y.L.); yinyitao@ihep.ac.cn (Y.Y.); max@ihep.ac.cn (X.M.); ychen@ihep.ac.cn (Y.C.); zhangsn@ihep.ac.cn (S.Z.)
\\
$^{2}$ \quad University of Chinese Academy of Sciences, Chinese Academy of Sciences, Beijing 100049, China\\
$^{3}$ \quad National Astronomical Observatory, Chinse Academy of Sciences, Beijing 100101, China; hengxu@bao.ac.cn (H.X.)\\
$^{4}$ \quad Department of Astronomy, School of Physics, Peking University, Beijing 100871, China; kjlee@pku.edu.cn (K.L.)\\
$^{5}$ \quad Xinjiang Astronomical Observatory, Chinese Academy of Sciences, Urumqi 830011, China; yuanjp@xao.ac.cn (J.Y.)\\
$^{6}$ \quad School of Information Engineering, Nanchang University, Nanchang 330031, China\\
}

\corres{Correspondence: zhengsj@ihep.ac.cn}

\abstract{Millisecond pulsars (MSPs) are known for their long-term stability. Using six years of observations from the Neutron Star Interior Composition Explorer (NICER), we have conducted an in-depth analysis of the X-ray timing results for six MSPs: PSRs B1937+21, B1821$-$24, J0437$-$4715, J0030+0451, J0218+4232,  and J2124$-$3358. The timing stability parameter $\sigma_z$ has been calculated, revealing remarkable timing precision on the order of $10^{-14}$
for PSRs B1937+21 and J0437$-$4715, and 
$10^{-13}$ for PSRs B1821$-$24, J0218+4232, and J0030+0451 over a timescale of 1000 days. These findings underscore the feasibility of autonomous in-orbit timekeeping using X-ray observations of MSPs. In addition, the consistency of long-term spin-down noise in the X-ray and radio bands has been investigated by comparison with IPTA radio data.}


\begin{document}

\section{Introduction}

Pulsars, particularly MSPs, exhibit long-term stability in their pulsation periods, which rivals that of current atomic timescales \citep{1991IEEEP..79.1054T,1997A&A...326..924M,2009MNRAS.400..951V,2010ApJ...725.1607S,2010MNRAS.402.1027H,2020MNRAS.491.5951H}. Consequently, high-precision pulsar timing can constrain the mass--radius relationship of neutron stars (NSs) and elucidate their emission mechanisms \citep{2019ApJ...887L..25B, 2019ApJ...887L..26B}. Moreover, it can facilitate the detection of nanohertz gravitational waves from supermassive black hole binaries well in advance of system coalescence \citep{2005ApJ...625L.123J, 2023RAA....23g5024X}, and enable precise measurement of solar system planetary masses~\citep{2010ApJ...720L.201C}. Furthermore, given the separate emission regions for radio and X-ray radiation, investigation of the consistency of spin-down noise in the X-ray and radio bands will underscore the understanding the origins of pulsar emissions, the characteristics of timing noise, and the evolutionary processes involved \citep{2010Sci...329..408L,2017ApJ...849...19G,2021ApJ...909..174D}.

Pulsars also play an important role in positioning, navigation, and timing (PNT). As the number of spacecraft launched into Earth's orbit, the vicinities of the Moon and Mars, and beyond into deep space continues to grow, the demand for PNT capabilities in space is becoming ever more critical. Current navigation and timekeeping for spacecraft in low-Earth orbits predominantly depend on Global Navigation Satellite Systems (GNSSs) and ground-based tracking, whereas deep-space missions rely primarily on radio technologies and space--Earth time comparisons. Pulsar-based navigation, a vital auxiliary solution, has been proposed and successfully tested in orbit \citep{2005PhDT........30S, 2017SSPMA..47i9505Z,2019JATIS...5a8003H,2019ApJS..244....1Z,2020AcAau.176..531Y,2023JGCD...46.1850w}. In addition, pulsars will offer the potential for timekeeping services as a ‘cosmic clock', compensating for biases in satellite-mounted atomic clocks or oscillators due to varying space environments, long-term drift, inadequate relativity corrections, or infrequent synchronization with Earth's high-stability clocks \citep{2021ApJ...908..184B}. Pulsars stand as prime candidates to provide an external absolute time standard, particularly for space missions in diverse orbits. 
A pulsar-based timescale, utilizing a pulsar timing array, has been proposed ,employing high-precision times of arrival (TOAs) detected by ground-based radio telescopes \citep{2020MNRAS.491.5951H}. X-ray timing offers unique advantages over radio due to its negligible dispersion measure (DM) effects, miniaturized detectors, and independence from ground-based systems, despite sharing similar pulsar parameters and intrinsic noise profiles with radio observations. Frequency steering of spaceborne clocks has been demonstrated based on XPNAV-1 observations \cite{2023CJA.36.266}. However, the accuracy of X-ray TOA detections is still a bottleneck, and a higher-precision demonstration still needs to be conducted.

The Neutron Star Interior Composition Explorer (NICER), with its unparalleled capability to detect millisecond pulsars, enables detailed analyses of the time-stability of X-ray pulses, assessing their utility for on-orbit timekeeping. The six X-ray MSPs, including PSRs B1937+21, B1821$-$24, J0437$-$4715, J0030+0451, J0218+4232, and J2124$-$3358, demonstrate high X-ray fluxes, small rotation periods, and minimal period derivatives, coupled with high stability (although PSR B1937+21 and B1821$-$24 show low-frequency timing irregularities dominate the timing residuals \cite{1994ApJ...428..713K,2009MNRAS.400..951V}), rendering them excellent candidates for timekeeping purposes. Extensive observational data of NICER for these six MSPs span over six years. Here, the new timing results of the MSPs with more than six years of observations from NICER have been obtained and demonstrate the feasibility of using pulsars for autonomous space navigation.

\section{Data Reduction and Timing Analysis}

\subsection{Observation}

NICER, an X-ray observatory affixed to a movable arm on the exterior of the International Space Station (ISS), has been operational since June 2017 \citep{2017NatAs...1..895G}. Designed specifically to investigate the X-ray emissions of NSs, NICER's primary objective is the high-precision timing of X-ray-emitting pulsars to constrain NSs' mass--radius relationships and to study their high-energy emission mechanisms. The exposures of PSRs B1937+21, B1821$-$24, J0437$-$4715, J0030+0451, J0218+4232, and J2124$-$3358 were around several mega-seconds over the period from June 2017 to September 2023, as listed in Table \ref{Tab1}, which could be utilized to supply more accurate timing results.

\begin{table}[H]\setlength{\tabcolsep}{4mm}
\caption{The observational information of the six MSPs.}
\label{Tab1}
\begin{adjustwidth}{-\extralength}{0 cm}
\centering
\small
\begin{tabular}{lllllll}
\toprule
\multirow{2}{*}{\textbf{Pulsar Name}} &	\multirow{2}{*}{\textbf{No. of Obs.}	}& \textbf{Exposure} & \textbf{Net Exposure} & \textbf{Count Rate \textsuperscript{1}} & \textbf{Energy} & \multirow{2}{*}{\textbf{Filtered Photons}}\\
 &	 &  \textbf{Ms}  &  \textbf{Ms}  & \textbf{Counts/s} & \textbf{keV}\\
\midrule
	PSR B1937+21	&    789	& 2.54 & 1.65  & 0.33 & 1.15--5.55 & 543,133\\
	PSR B1821$-$24	&    598	& 1.47 & 0.67  & 0.61 & 1.0--5.5  & 407,839\\
	PSR J0437$-$4715	&    833	& 2.93 &  2.50 & 1.87 & 0.25--1.85 & 4,681,958\\
	PSR J0030+0451	&    731	& 3.98 & 3.70  & 0.72 & 0.25--1.45 & 2,679,284\\
	PSR J0218+4232	&    675	& 2.74 & 1.85  & 0.44 & 0.8--6.2 & 819,464\\
	PSR J2124$-$3358	&    630	& 2.23 & 2.09  & 0.49 & 0.25--1.59 & 1,018,017\\
\bottomrule
\end{tabular}
\end{adjustwidth}
	\noindent{\footnotesize{$^{1}$ These count rates include both source and background contributions.}}
\end{table}\vspace{-13pt}

\subsection{Data Reduction}

Our data selection criteria are derived from those established by \citet{2019ApJ...874..160D,2019ApJ...887L..25B}. In order to increase the signal to noise ratio, we have applied additional data screening as follows: 
(1) Exclusion of ``hot'' detectors: Detector IDs 14, 34, and 54 have been consistently excluded across all datasets due to their anomalously high count rates. (2) Background filtering: Photon events from periods with significant background interference are removed. For PSRs B1937+21, B1821$-$24, and J0218+4232 we have excluded 8-second intervals where the average count rate exceeds 2 counts per second from the Good Time Intervals (GTIs), in accordance with \citet{2019ApJ...874..160D}. For PSRs J0437$-$4715, J0030+0451, and J2124$-$3358 we have removed 16-second intervals where the average count rate in the 0.25--8\,keV band exceeds 3.5, 3.0, and 2.8\,counts/s, as per the recommendations of \citet{2019ApJ...887L..25B}. (4) Energy band optimization: To achieve high-precision ToA measurements, the energy bands have been carefully selected for each pulsar to maximize the significance of the pulsed signal as suggested \citep{2019ApJ...874..160D, 2019ApJ...887L..25B}. The bands are 1.15--5.55 keV for PSR B1937+21, 1.0--5.5 keV for PSR B1821$-$24, 0.8--6.2 keV for PSR J0218+4232, 0.25--1.85 keV for PSR J0437$-$4715, 0.25--1.45 keV for PSR J0030+0451, and 0.25--1.59 keV for PSR J2124$-$3358.

The events from the six pulsars have been successfully filtered with the above criteria. For PSR B1937+21, a total of 543,133 photons meet our selection criteria across 1.65\,Ms of net exposure time, yielding an average count rate of 0.33 counts/s. In the case of PSR B1821$-$24, we have identified 407,839 photons over 0.67\,Ms of clean exposure, with an average count rate of 0.61\,counts/s. For PSR J0218+4232, the data comprise 819,464 photons across 1.85\,Ms of net exposure, averaging 0.44\, counts/s. PSR J0437$-$4715 yielded 4,681,958 photons over 2.50\,Ms of clean exposure, resulting in an average count rate of 1.87\,counts/s. For PSR J0030+0451, we obtained 2,679,284 photons over 3.70\,Ms of clean exposure, with an average count rate of 0.72\,counts/s. Lastly, PSR J2124$-$3358 contributed 1,018,017 photons over 2.09\,Ms of clean exposure, averaging 0.49\,counts/s. It is important to note that these count rates include both source and background contributions.

\subsection{ToA Calculation}

To obtain ToAs of each pulsar, the arrival time of each photon at the local observatory is corrected to the solar system barycenter (SSB) and then folded with the pulsar's ephemeris. The initial pulsar ephemerides are sourced from the International Pulsar Timing Array (IPTA) Data Release 2 (DR2) \cite{2019MNRAS.490.4666P} and refined with NICER data \cite{2012ApJS..199...32G}. Owing to decades of monitoring, the positions, proper motions, parallaxes, and binary orbits of these pulsars have been determined with high precision by radio telescopes. Consequently, we fix these parameters in our analysis. The rotational period and its derivative serve as initial inputs and are updated with the X-ray observations. Our timing analysis employs the DE436 JPL ephemeris, Barycentric Dynamical Time (TDB) units, and the TT(BIPM2022) clock realization. Additionally, we have incorporated the GNU Multiple Precision Arithmetic Library (GMP) (\url{https://gmplib.org/} (accessed on 13, December, 2023)) to facilitate high-precision data recording at the 1~ns~level.

For each pulsar, we have constructed a high-precision ``standard'' pulse profile using six years of data. Due to the substantial variation in photon counts across different observation IDs (ObsIDs)—with some containing fewer than 100 photons—we have segmented the entire dataset based on count rates. Each segment comprises approximately 50,000 to 200,000 photons within a 30-day span. We obtain pulsed profiles for each segment and calculate ToAs through cross-correlation analysis with the standard profile. The ToA error is calculated by performing Gaussian sampling across the pulse profile. The pulsar period parameters and timing residuals are then refined using Tempo2 \citep{2006MNRAS.369..655H, 2006MNRAS.372.1549E}. 

\section{Timing Results} \label{sec:timing}

\begin{figure}[H]
\includegraphics[width=10.5cm]{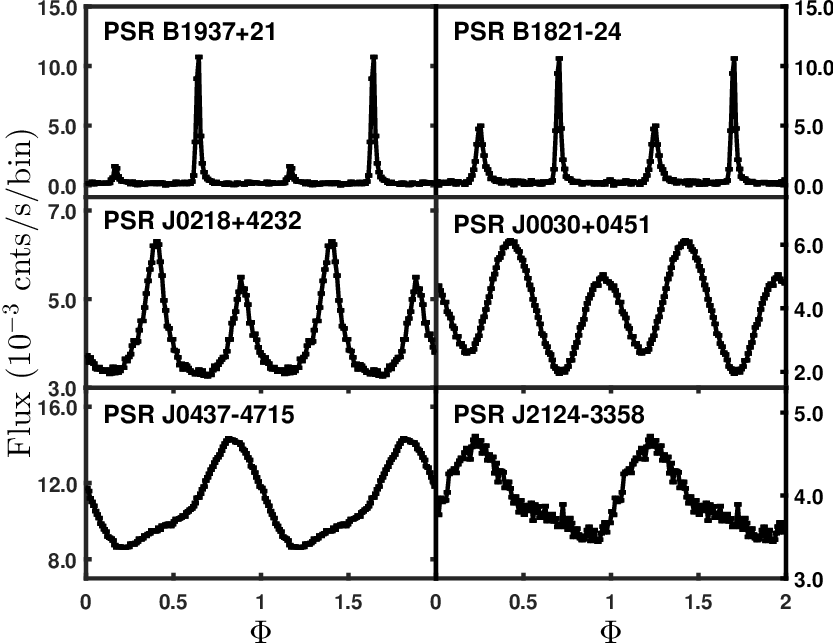}
\caption{The pulse profiles of PSRs B1937+21, B1821$-$24, J0437$-$4715, J0030+0451, J0218+4232, and J2124$-$3358 as a \textls[-25]{function of their phases ($\phi$). The flux in each bin is calculated with background~subtracted.} \label{fig:profile_all}}
\end{figure}

The frequency and its first-order derivative in each pulsar ephemeris have been updated by refitting all ToAs, while the other parameters remain fixed as in IPTA. The higher-quality pulse profiles of the six pulsars are shown in Figure \ref{fig:profile_all}. With the six years of observational data, we have been able to obtain a highly accurate pulse profile. PSRs B1937+21, B1821$-$24, and J0218+4232 consist of a main pulse and an interpulse approximately $180$ degrees apart in phase. For PSRs J0437$-$4715, J0030+0451, and J2124$-$3358, the profiles exhibit an approximate sinusoidal shape, indicative of radiation primarily from one or more hot spots on the  surface of the neutron star. Additionally, the asymmetric pulse profiles and the distinct ``hump'' observed in PSRs J0437$-$4715 and J2124$-$3358 are attributed to a secondary hot spot \cite{2013ApJ...762...96B}.

\begin{figure}[H]
\includegraphics[width=12.5cm]{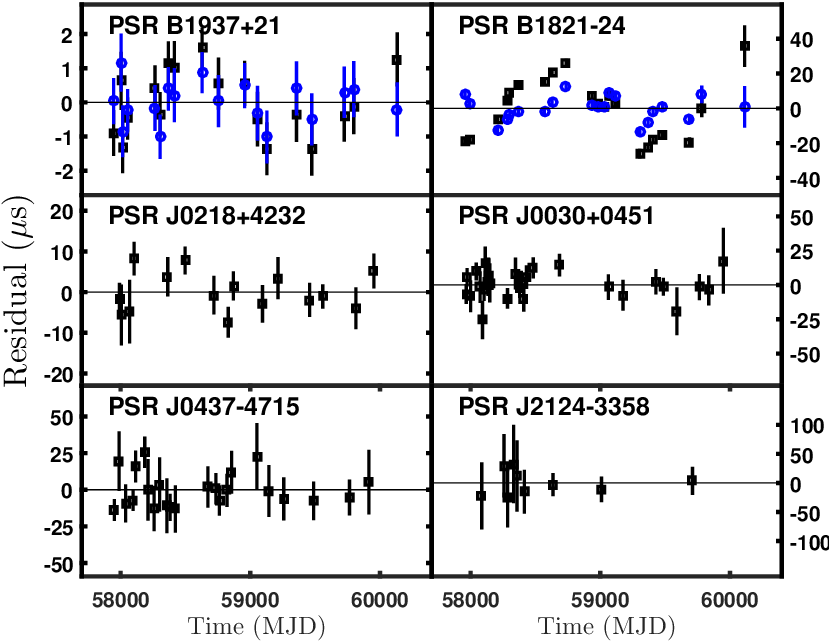}
\caption{The X-ray timing residuals of six pulsars obtained from NICER observations as a function of modified Julian date (MJD). The black squares in the six panels represent the timing residuals for PSRs B1937+21, B1821$-$24, J0437$-$4715, J0030+0451, J0218+4232, and J2124$-$3358. These residuals are calculated from the updated pulsar ephemerides, considering the frequency and first-order derivatives of frequency, as listed in Table \ref{Tab2}. For PSRs B1937+21 and B1824$-$24, the blue circles show the timing residuals with second-order derivatives of frequency included.
\label{fig:resi_all}}
\end{figure}

The timing residuals obtained with the updated pulsar ephemerides may reflect various factors, including pulsar timing noise, the gravitational wave background, clock errors, solar system ephemeris and pulsar ephemeris errors, and other uncalibrated instrumental effects. The dispersion measure (DM) and its variation have a negligible impact for X-ray observation as mentioned above. The gravitational wave background is also considered negligible \cite{2023RAA....23g5024X}. The event times have been calibrated and corrected with a precision of about 40\,ns and are referenced to UTC with an accuracy better than 100\,ns \cite{2022ApJ...928...67H}. The influence of the solar system ephemeris is expected to reach about 100\,ns over typical periods of 10\,years \cite{2020ApJ...893..112V}. Furthermore, inaccuracies in the IPTA radio pulsar ephemerides arise from covariances between certain parameters and variations in the DM, which could potentially introduce unaccounted-for errors in our timing measurement. Consequently, the timing residuals of the six pulsars, containing all the factors mentioned above, are shown in Figure \ref{fig:resi_all}, ranging from a few microseconds to tens of microseconds. For PSR B1937+21, we observe the lowest X-ray band residuals, with the RMS within 1\,$\upmu s$ due to the smallest spin period and the sharpest (and narrowest) profile, as illustrated in Figure \ref{fig:resi_all}. Therefore, PSR B1937+21 holds significant potential for applications in X-ray timekeeping. Low-frequency timing irregularities have been observed over a period of more than three years \cite{2009MNRAS.400..951V}. However, if these timing irregularities can be accurately modeled and removed, it will raise the possibility of producing an essentially stable X-ray clock \cite{2010Sci...329..408L}. 

The second derivative of the spin frequency, $\ddot{\nu}$, which may have intrinsic or kinematic origins in MSPs, is an important consideration for high-precision pulsar timing. Therefore, we have refitted the ephemerides of the six pulsars, including the second derivatives of frequency, using TEMPO2. A statistically significant non-zero second derivative of the frequency (with a significance greater than 3$\sigma$) is detected only in PSRs B1937+21 and B1821$-$24. As shown in Figure \ref{fig:resi_all}, the cubic structures in the timing residuals are suppressed effectively with $\ddot{\nu}$ included, especially for PSR B1821$-$24. The root mean square of the timing residuals of PSR B1937+21 decreases from 0.9\,$\upmu s$ to 0.6\,$\upmu s$ while it decreases from 16.7\,$\upmu s$ to 6.7\,$\upmu s$ for PSR B1821$-$24.

For PSR B1937+21, $\ddot{\nu}$ is obtained with $(-1.8 \pm 0.4)\times 10^{-26}$\,$\rm{Hz\,s^{-2}}$ with high significance, which is somewhat larger, but consistent within three standard deviations, from \cite{2019MNRAS.490.4666P}, while quite different from \cite{2009PhDT.......100V}. New physical explanations are required, as neither the spin-down due to the braking process nor a distant binary companion can account for the statistically significant $\ddot{\nu}$ \cite{2019MNRAS.490.4666P}. For PSR B1821$-$24, constraints on $\ddot{\nu}$ have varied significantly across studies, with reported values of $(29.42 \pm 15.75)\times 10^{-27}$, $(-175 \pm 7) \times 10^{-27}$,  $(-26.49 \pm 0.05) \times 10^{-27}$, or $(200 \pm 540) \times 10^{-27}\,\rm{Hz\,s^{-2}}$ \citep{2019MNRAS.490.4666P,1996A&A...311..179C,2013ApJ...778..106J, 2009PhDT.......100V}. Our findings yield $\ddot{\nu}=(-2.1 \pm 0.2)\times 10^{-25}\,\rm{Hz\,s^{-2}}$, which is consistent with the results of \citep{1996A&A...311..179C,2009PhDT.......100V} considering the errors. However, it is worth mentioning that $\ddot{\nu}$ is not a good measure of stability anyway, since it shows wildly different values depending on the range of data, as demonstrated by \cite{2010MNRAS.402.1027H}. 


\vspace{-8pt}

\begin{table}[H]\setlength{\tabcolsep}{2.6mm}
\caption{The timing parameters of the six MSPs.}
\label{Tab2}
\small
\begin{tabular}{llllll}
\toprule
\multirow{2}{*}{\textbf{Pulsar Name}} & \textbf{PEPOCH} & \boldmath{$\nu$} & \boldmath{$\dot{\nu}$}  & \boldmath{$\ddot{\nu}$}  & \textbf{RMS} \\
 & \textbf{(MJD)} & \textbf{(Hz)} &  \textbf{(}\boldmath{$10^{-15}\,{\rm Hz\,s^{-1}}$}\textbf{) }&  \textbf{(}\boldmath{$10^{-25}\,{\rm Hz\,s^{-2}}$}\textbf{)} & \textbf{(}\boldmath{$\upmu s$}\textbf{)} \\
\midrule
B1937+21 &	57,920 & 641.92822360944(1) & $-$43.3081(1) &	$-$   & 0.9 \\
 &	 & 641.92822360938(1) & $-$43.3064(4) &$-$0.18(4)    & 0.6 \\  
B1821$-$24 &	57,920 & 327.40554427924(9) & $-$173.511(1) &     & 16.7 \\
 &	 & 327.40554427917(9) & $-$173.510(2) &	$-$2.1(2))    & 6.7 \\
J0437$-$4715 &	57,920 & 173.68794799450(3) & $-$1.7277(4)  &$-$&11.6 \\
J0030+0451 &	57,920 & 205.53069901678(4) & $-$0.4292(4)  &$-$& 9.9\\
J0218+4232 &	57,920 & 430.46105760208(3) & $-$14.3425(3)  &$-$& 4.6\\
J2124$-$3358 &	57,920 & 202.7938966770(2) & $-$0.846(2)  &$-$& 6.7 \\
\bottomrule
\end{tabular}
\end{table}

\section{Evaluation of Rotational Stability} \label{sec:stability}

Clock stability is typically described by the Allan variance, $\sigma_y^2$, which is designed to quantify instability in 
clocks operating at a constant frequency. However, pulsars are known to have a frequency derivative due to the gradual slowdown of their rotation. According to the timing analysis, we can remove the effects of pulsar frequency, frequency derivative, proper motion, parallax, and other systematic influences. The remaining timing residuals include the measurement error, uncalibrated instrumental error, and intrinsic timing noise, the latter providing insight into the rotational stability of the pulsar and determining its potential as a precise timekeeping clock. To evaluate the rotational stability, another parameter, $\sigma_z^2$, has been introduced to evaluate the rotational stability by calculating the variations in the timing residuals \cite{1997A&A...326..924M}.

For an interval of length $\tau$ starting at time $t_0$, we can fit a cubic polynomial to the timing residuals in that interval,
$$X(t)=c_0+c_1(t-t_0)+c_2(t-t_0)^2+c_3(t-t_0)^3$$
where $X(t)$ minimizes the sum of $[(x_i-X(t_i))/\sigma_i]^2$ over all TOA $t_i$ with residuals $t_i$ and uncertainties $\sigma_i$ . Then, 
$$\sigma_z=\frac{\tau^2}{2\sqrt{5}}<c_3^2>^{1/2}$$
where angle brackets denote the weighted average over the third-order coefficients of the best-fit polynomials of all
non-overlapping intervals of length $\tau$ within the set of timing residuals. 

\begin{figure}[H]
\includegraphics[width=10cm]{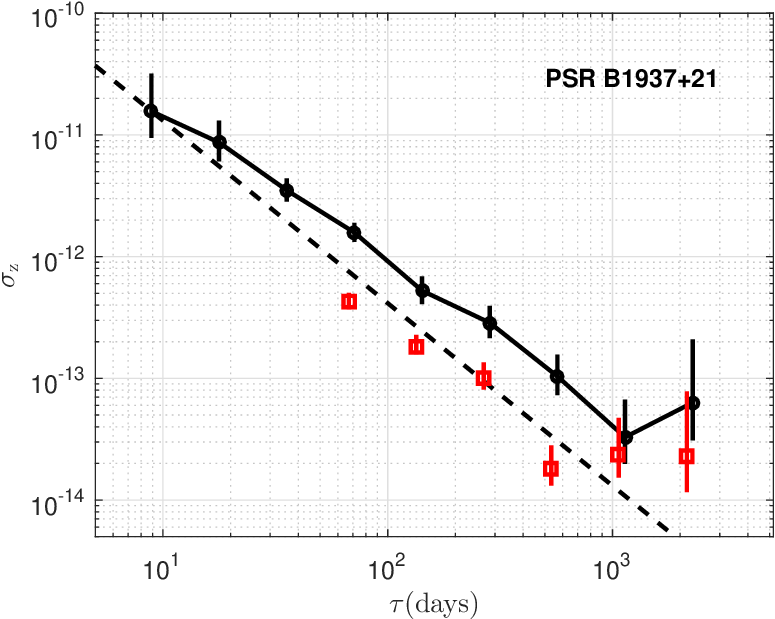}
\caption{Measure of timing stability, $\sigma_z$, vs. the data span over which a third-order polynomial is fitted to the timing residuals of PSR B1937+21. {Black circle points show $\sigma_z$ for NICER data, and red square points show the radio results from \cite{2019ApJ...874..160D}}. A dashed line shows the slope for the case where timing precision is limited by white noise only.
\label{fig:sigma_z_1937}}
\end{figure}
\begin{figure}[H]
\includegraphics[width=10cm]{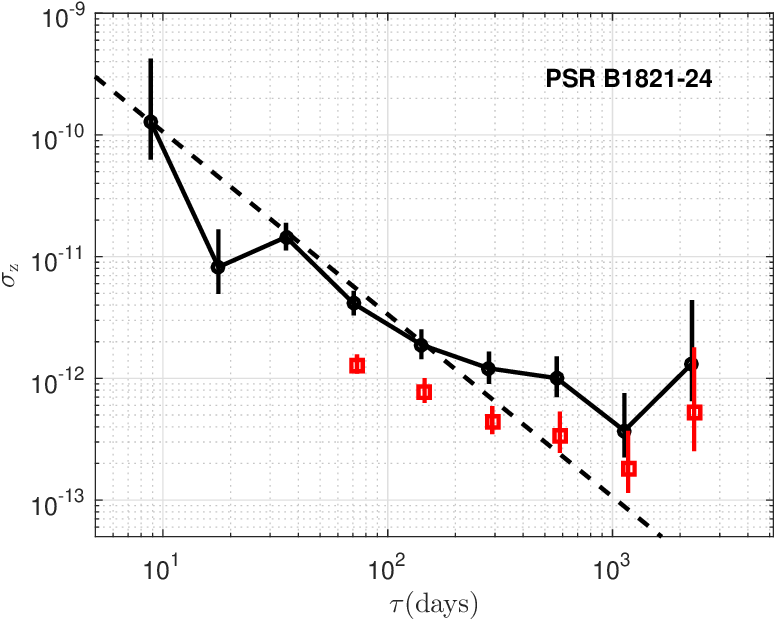}
\caption{Same as Figure \ref{fig:sigma_z_1937} but for PSR B1821$-$24. 
\label{fig:sigma_z_1821}}
\end{figure}
\begin{figure}[H]
\includegraphics[width=12.5cm]{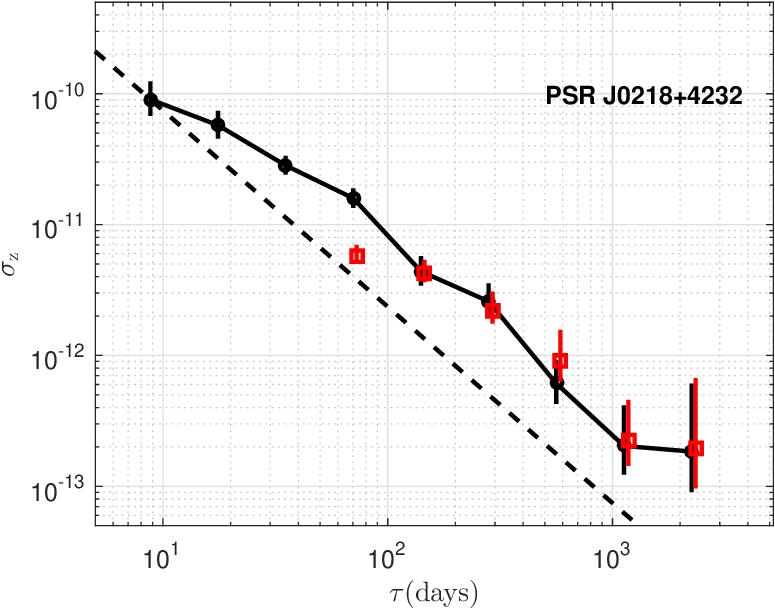}
\caption{Same as Figure \ref{fig:sigma_z_1937} but for PSR J0218+4232. 
\label{fig:sigma_z_0218}}
\end{figure}
\begin{figure}[H]
\includegraphics[width=12.5cm]{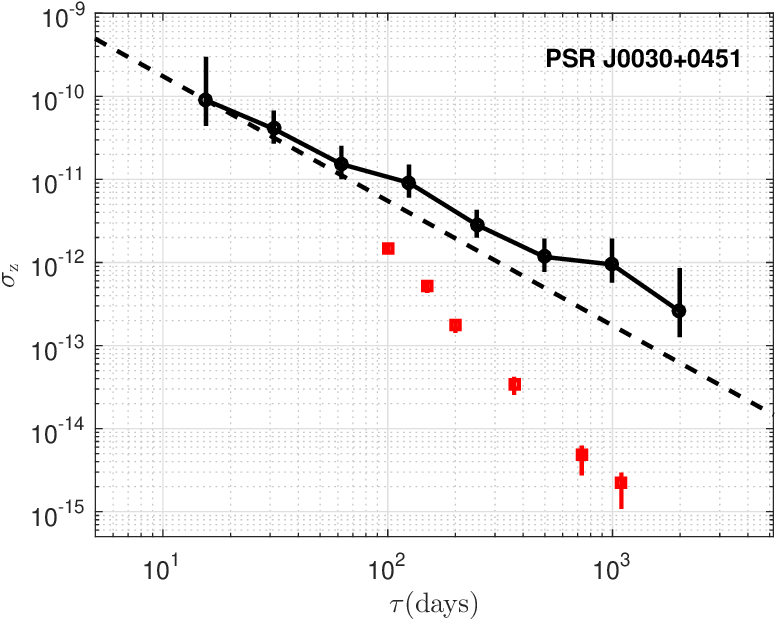}
\caption{Same as Figure \ref{fig:sigma_z_1937} but for PSR J0030+0451. {The red square points show the radio results with IPTA DR2 dataset.}
\label{fig:sigma_z_0030}}
\end{figure}
\begin{figure}[H]
\includegraphics[width=12.5cm]{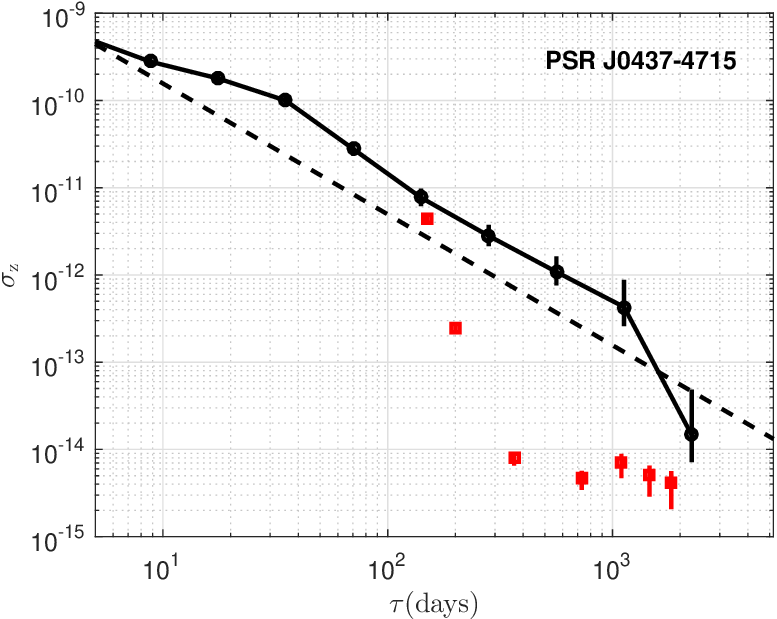}
\caption{Same as Figure \ref{fig:sigma_z_1937} but for PSR J0437$-$4715. {The red square points show the radio results with IPTA DR2 dataset.}
\label{fig:sigma_z_0437}}
\end{figure}
\begin{figure}[H]
\includegraphics[width=12.5cm]{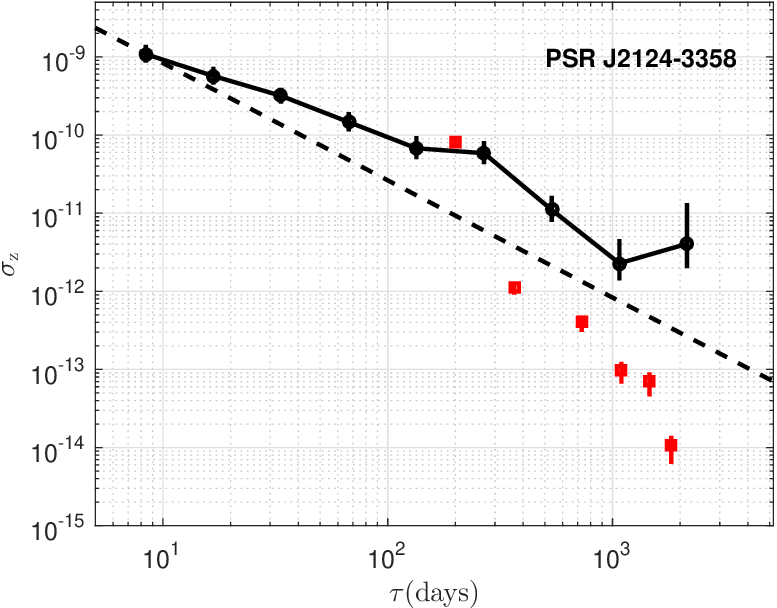}
\caption{Same as Figure \ref{fig:sigma_z_1937} but for PSR J2124$-$3358. {The red square points show the radio results with IPTA DR2 dataset.}
\label{fig:sigma_z_2124}}
\end{figure}
\begin{figure}[H]
\includegraphics[width=12.5cm]{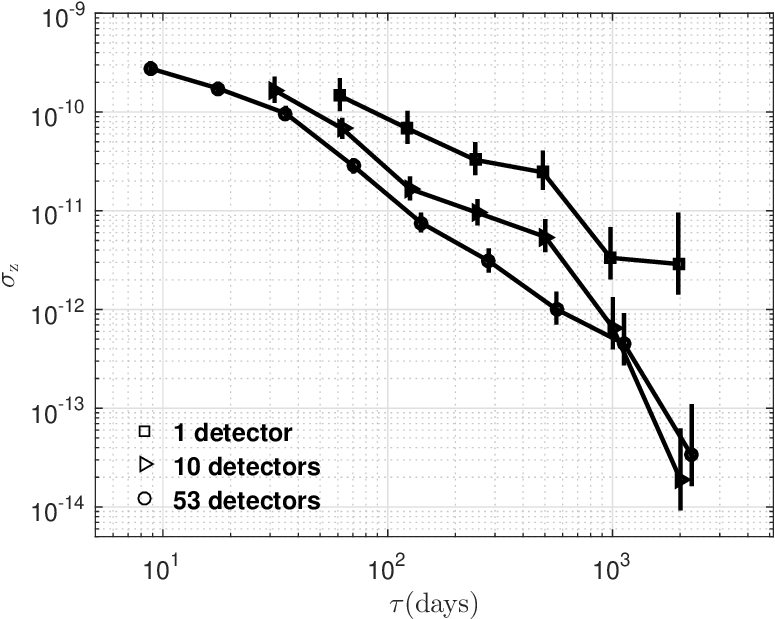}
\caption{Same as Figure \ref{fig:sigma_z_1937} but for PSR J0437$-$4715. A varying number of detectors from NICER were employed in the analysis: black squares represent data from a single detector, black triangles correspond to data from ten detectors, and black circles indicate data from all but three of the hottest detectors, totaling fifty-three.
\label{fig:sigma_z_0437_det}}
\end{figure}

The timing stability, $\sigma_z$, is plotted as a function of the data span, $\tau$, for the six millisecond pulsars and is presented in \Crefrange{fig:sigma_z_1937}{fig:sigma_z_2124}. For a direct comparative analysis, the results are juxtaposed with observations from the radio band. Furthermore, in the scenario where the residuals are solely influenced by white noise, we expect $\sigma_z$ to scale with $\tau$ as $\sigma_z \propto \tau^{-1.5}$, as described by \cite{1997A&A...326..924M}. This relationship is depicted by the dashed lines included in each of the figures for reference. Despite some deviation at lower time spans, it demonstrate comparable stability in both the radio and X-ray bands for PSRs B1937+21, B1821$-$24, and J0218+4232, showing the consistence of the timing between the X-ray and radio bands. However, for PSRs J0030+0451, J0437$-$4715, and J2124$-$3358, the stability indices in the X-ray band are lower than those in the radio band, with the largest discrepancies exceeding a factor of 100 {\citep{2019ApJ...874..160D,2011RvMP...83....1H}}. As shown in Figure \ref{fig:resi_all}, the time residuals for the latter three pulsars are larger due to the broad and sinusoidal-like profiles, {which suggests that the timing stability is likely still limited by white noise. However, other possible explanations corresponding to the X-ray observations will be considered.} Consequently, a varying number of detectors from NICER were utilized in the analysis to evaluate the timing stability in relation to different detector areas. As illustrated in Figure \ref{fig:sigma_z_0437_det}, an increase in detector area leads to enhanced timing stability, thereby reinforcing the predominance of white noise. Consequently, for PSRs J0030+0451, J0437$-$4715, and J2124$-$3358, a larger detection area, more observation time, and reduced background noise are required to suppress white noise and achieve results comparable to those in the radio band.


PSR B1937+21 emerges as an excellent candidate for precise timekeeping, approaching $10^{-14}$ for data spans around 1000\,days, consistent with radio timing results \citep{2009MNRAS.400..951V,2019ApJ...874..160D}. However, the red noise is also very remarkable. In addition, the reported upturn at $\geq$500\,days for PSR B1821$-$24 is confirmed in detail \citep{2019ApJ...874..160D}. The upturn observed at $\geq$1000 days, also consistent with radio datasets for PSRs B1937+21, B1821$-$24, and J0218+4232 (as shown in \citet{2019ApJ...874..160D}), indicates a cutoff of stability over longer timescales. While for PSRs J0030+0451, J0437$-$4715, and J2124$-$3358, the stability is still dominated by white noise. 

\vspace{-8pt}

\vspace{-6pt}

\vspace{-6pt}

\section{Summary and Conclusions} \label{sec:summary}

X-ray pulsar-based timekeeping presents a promising application for space missions, with the long-term stability of millisecond X-ray pulsars (MSPs) being a critical parameter. Utilizing  six years of observations from NICER, the timing stability, $\sigma_z$, has been assessed for six candidate pulsars: PSRs B1937+21, B1821$-$24, J0437$-$4715, J0030+0451, J0218+4232, and J2124$-$3358. Despite the presence of red noise and/or white noise, long timing stability of $10^{-14}$ for PSRs B1937+21 and J0437$-$4715, and 
$10^{-13}$ for PSRs B1821$-$24, J0218+4232, and J0030+0451 has been achieved over a span of 1000 days, supporting the feasibility of timekeeping. It is important to note that a thorough investigation of the red noise and efforts to mitigate or minimize its impact on timing residuals will further enhance the utility of this approach. Moreover, by comparison with IPTA radio data, the consistency of long-term spin-down noise in the X-ray and radio bands has been investigated. It shows a similar upturn for PSRs B1937+21, B1821-24, and J0218+4232, which could provide meaningful information on the emission and spin-down mechanisms of neutron stars.

\vspace{6pt} 




\authorcontributions{Conceptualization, S.Z. (Shijie Zheng), D.H., and M.G.; methodology, S.Z. (Shijie Zheng), H.X., K.L., J.Y., and L.Z.; software, H.W.; validation, Y.L. and Y.Y.; formal analysis, X.M.; writing---original draft preparation, S.Z. (Shijie Zheng) and D.H.; writing---review and editing, M.G., Y.C., and S.Z. (Shuangnan Zhang). All authors have read and agreed to the published version of the~manuscript.}

\funding{This work is supported by the National Key R\&D Program of China (2022YFF0711404, 2021YFA0718500), and the National Natural Science Foundation of China (Grant No. 12333007). Heng Xu is supported by the Project funded by China Postdoctoral Science Foundation No. 2023M743518, Kejia Lee and Heng Xu are supported by National SKA Program of China 2020SKA0120100.}

\dataavailability{All the data used in this paper can be found at NASA's HEASARC website, https://heasarc.gsfc.nasa.gov/cgi-bin/W3Browse/w3browse.pl.} 

\acknowledgments{We would like to appreciate the NICER mission and the High Energy Astrophysics Science Archive Research Center (HEASARC) for the good observation data and International Pulsar Timing Array for the pulsar ephemerides.} 

\conflictsofinterest{The authors declare no conflicts of interest.} 



\abbreviations{Abbreviations}{
The following abbreviations are used in this manuscript:\\

\noindent 
\begin{tabular}{@{}ll}
MSP & Millisecond pulsars\\
NS & Neutron star\\
PNT & Position, navigation, and timing\\
GNSS & Global Navigation Satellite System\\
TOA & Times of arrival\\
NICER & Neutron Star Interior Composition Explorer\\
GTI & Good Time Interval\\
IPTA &  International Pulsar Timing Array
\end{tabular}
}

\begin{adjustwidth}{-\extralength}{0cm}

\reftitle{References}
\PublishersNote{}
\end{adjustwidth}
\end{document}